\documentstyle[a4wide,epsf,11pt,titlepage]{article}

\newcounter{nref}
\newcommand{\bbib}{%
  \renewcommand{\refname}{\large\bf References}%
  \begin{thebibliography}{99}%
  \setcounter{enumiv}{\thenref}}
\newcommand{\ebib}{%
  \end{thebibliography}%
  \setcounter{nref}{\arabic{enumiv}}}
\newcommand{\head}[3]{%
  \setcounter{nref}{0}%
  \thispagestyle{empty}%
  \section*{\LARGE\bf #1}%
  \stepcounter{section}%
  \addcontentsline{toc}{section}{#1}%
  \large\itshape%
  #2\\\vspace{0.1pt}\\%
  #3%
  \normalsize\upshape%
  \bigskip}

\begin{document}


\head{Explosive Nucleosynthesis: Coupling Reaction Networks to 
      AMR Hydrodynamics}
     {K.\ Kifonidis$^1$, T. \ Plewa$^{2,1}$, E. \ M\"uller$^1$}
     {$^1$ Max-Planck-Institut f\"ur Astrophysik,
      Karl-Schwarzschild-Strasse 1, D-85740 Garching, Germany\\
      $^2$ Nicolaus Copernicus Astronomical Center, Bartycka 18, 
      00716 Warsaw, Poland}


\subsection{Introduction}

Observations of SN 1987A revealed that extensive mixing had taken
place in the exploding envelope of the progenitor Sk~-69~202.  
Especially the early detection of X and $\gamma$-rays  
\cite{Kifonidis.Dotani}, \cite{Kifonidis.Matz}, the broad profiles 
of infrared Fe\,II and Co\,II lines \cite{Kifonidis.Colgan}, 
\cite{Kifonidis.Haas} as well as
modelling of the light curve \cite{Kifonidis.Arnett},
\cite{Kifonidis.Woosley} indicated that $\rm ^{56}Ni$ was mixed from
the layers  close to the collapsed core, where it was explosively
synthesized, out to the hydrogen envelope where the highest expansion
velocities  occurred. 

Multidimensional hydrodynamic models of the late phases of the
explosion (starting several minutes after core bounce) while
successful in confirming that mixing due to Rayleigh-Taylor
instabilities did indeed occur after the explosion shock had passed
the C,O/He and He/H interfaces, have hitherto failed to yield the
amount of mixing observed \cite{Kifonidis.FMA}, \cite{Kifonidis.HB91},
\cite{Kifonidis.MFA}. However, Herant and Benz \cite{Kifonidis.HB92}
have shown that velocities in line with the observations could be
obtained if one {\em artificially\/} mixed $\rm ^{56}Ni$ in the very
early phases of the explosion out to layers which later suffer
from the Rayleigh-Taylor instabilities.

In the light of results from recent multidimensional simulations of
the (neutrino driven) explosion mechanism itself which revealed large
scale anisotropies, mixing and overturn due to convective motions
taking place within about one second after core bounce behind the
revived supernova shock, it has been argued \cite{Kifonidis.HBC},
\cite{Kifonidis.JM} that a physically satisfactory mechanism has been
found which might lead to the required amount of ``premixing'' and
thus resolve the nickel problem.  However, only very preliminary
multidimensional computations exist to date which attempt to follow
the mixing of nickel from the moment of nucleosynthesis until it
appears in the hydrogen envelope of the exploding star
\cite{Kifonidis.Nagataki}. Despite constant growth in computer
resources and steady advances in numerical algorithms such simulations
still pose a formidable task due to the large range of spatial and
temporal scales which have to be resolved.  Therefore, most of the
computations hitherto performed started from artificial {\em
spherical\/} models of the explosion itself.

In recent years the technique of Adaptive Mesh Refinement (AMR)
has been applied to several astrophysical problems 
(cf.~\cite{Kifonidis.Plewa}, 
\cite{Kifonidis.Walder}) and should allow
a consistent modelling of the complete evolution 
in two dimensions. In this contribution we address some of the 
computational difficulties encountered when trying to apply AMR 
to explosive nucleosynthesis and supernova envelope ejection.

\subsection{Adaptive Mesh Refinement}

\begin{figure}[t]
\centerline{\epsfxsize=0.7 \textwidth\epsffile{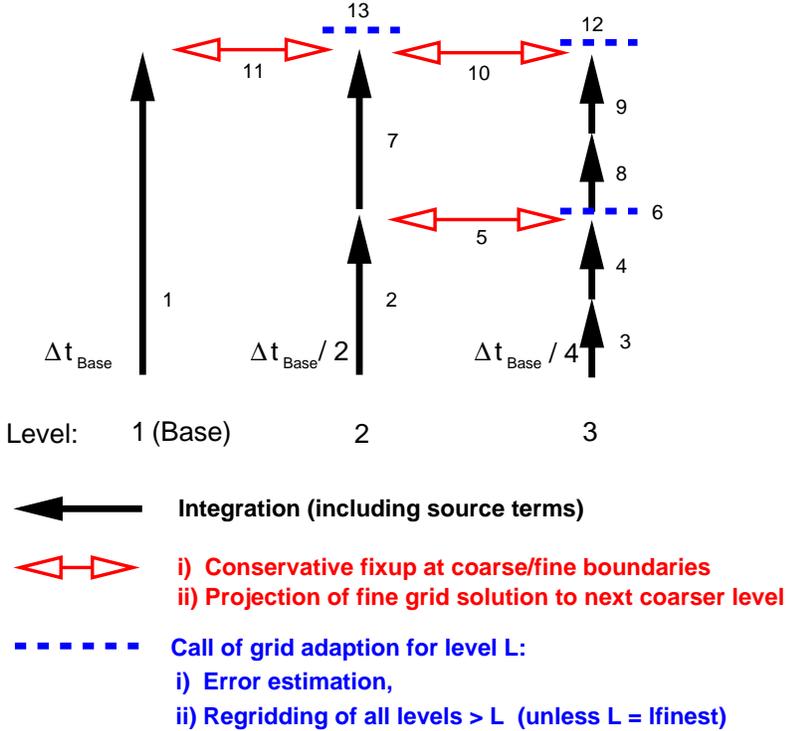}}
\caption{Integration of the grid hierarchy over a single base level time
step for 3 levels of refinement with a constant refinement factor 
$r = 2$. Note that grids at level $l+1$ have to be evolved with 
time steps $\Delta t_l /r$. The numbers indicate the actual sequence
of operations to be carried out. A regridding frequency of $K = 2$
was chosen in this example.}
\label{Kifonidis.fig1}
\end{figure}

AMR is an algorithm for the efficient solution 
of systems of time-dependent, hyperbolic partial differential equations
\cite{Kifonidis.BO}. An extended version of the basic AMR algorithm 
applied to the Euler equations of ideal, compressible flows 
has been discussed in \cite{Kifonidis.BC}. 
In essence, AMR provides a way to automatically adjust the
computational grid resulting from the discretization of 
the differential equations subject to the estimated error of the 
solution. Since in many cases this error is large only in some 
regions of the computational domain AMR usually offers large 
savings in CPU time and memory usage.

The AMR algorithm constructs and continuously updates a tree of nested
grid meshes or patches located on different {\em levels\/} in the tree
hierarchy.  Each level can be formed out of one or more patches
with the resolution changing between levels from lower (coarse) to
higher (fine) levels by arbitrary (but integer) factors in each
dimension. Patches forming a single level may partially overlap each
other or may cover distinct regions of the computational domain, but
those belonging to different levels must necessarily be ``properly
nested'', i.e.\  patches on a given level must be totally covered
by one or more patches located on the next coarser level.

Integration of the grids proceeds starting from the base level grid of
the lowest resolution, which covers the entire computational domain, and
recursively continues through the higher levels of the grid hierarchy
(Fig.~\ref{Kifonidis.fig1}). Some amount of communication between
the different levels is needed in order to obtain a consistent
solution. This includes averaging of the solution obtained on fine
patches and its projection down to parent patches.  Furthermore,
special attention is required at boundaries separating coarse and fine
grid cells.  The integration of fine grids is carried out using
boundary (ghost) zones which might have to be initialized by
interpolating data from coarser levels.  In the general case, numerical
fluxes calculated with higher resolution will differ from fluxes
calculated with lower resolution.  To ensure global conservation a 
correction pass over all coarser grid cells abutting fine grid cells 
is needed once both grid levels have been integrated to the 
same time. We refer the reader to \cite{Kifonidis.BC} for a more
detailed description of this procedure.

Finally, every $K$ time steps on a given level an error estimation
procedure is invoked, which yields an estimate of the local truncation
error. The regions where this value exceeds some predefined threshold,
$\epsilon$, are marked and later covered with new grid patches of
higher resolution.  Thereby flow features requiring high resolution
like shocks, contact discontinuities or strong gradients in the
solution are always followed with the higher level grids while regions
where the flow is essentially smooth are calculated at lower
resolution.  It is important to note in this context that newly
created fine grids might have to be initialized with data obtained by
interpolation from underlying coarser grids. As we will show below
this procedure may lead to serious numerical problems especially for
multi-component flows.

\subsection{Numerical tests}

In our numerical investigations we considered the problem of a
supernova explosion for a 15\,$\rm M_{\odot}$ model progenitor of
Woosley, Pinto and Ensman \cite{Kifonidis.WPE} in one dimension
assuming spherical symmetry and using the {\sc amra} code
\cite{Kifonidis.AMRA}. The hydrodynamic equations were solved with 
the direct Eulerian version of the Piecewise Parabolic Method
(PPM) as implemented in the {\sc prometheus} code \cite{Kifonidis.FMA}
although {\sc amra} can be used in conjunction with any hydrodynamic
scheme.

After removing the model's iron core the explosion was initiated by 
depositing an energy of $10^{51}$ ergs in form of a thermal bomb 
into the innermost region of the silicon shell.
We used five levels of refinement, with 256 zones on the base
grid (level 1) and refinement factors of 2, 4, 6, and 8 for patches on
levels 2, 3, 4, 5 respectively. This gave us an effective resolution
of 98\,304 equidistant zones.  The computational domain extended from
$1.4\times 10^8$\,cm up to $3.8\times 10^{11}$\,cm and covered about
the inner 1/10\,th of the star. Besides $\rm ^1H$, the 13
$\alpha$-nuclei from $\rm ^4He$ to $\rm ^{56}Ni$ were included.  A
realistic equation of state was used that contained contributions from
all considered nuclei as well as electrons, photons and
$e^+$/$e^-$-pairs. Gravity was taken into account and included the
contribution from the collapsed central core as well as self-gravity
of the envelope. The code was optimized to run efficiently on CRAY
shared memory systems.

The solution of the coupled system of hydrodynamic and nuclear rate
equations neccessitates a detailed description of the chemical
composition within the hydrodynamic scheme.  In {\sc prometheus} this
is achieved by solving additional continuity equations for each fluid
component, with the partial densities, $\rho X_{i}$, (where $X_i$
denotes the mass fraction of species $i$) as state variables. This
extension of basic PPM is reflected within {\sc amra} in two
ways. Firstly, the fixup procedure for fluxes at fine-coarse
boundaries is done for the partial densities in a similar way as for
the other conserved quantities. Secondly, fractional masses are
interpolated conservatively when boundary data for fine patches are
needed or when the hydrodynamic state for the interior of a newly
created fine patch has to be provided. Both steps may lead to serious
numerical problems due to the fact that the interpolation scheme does
{\em not\/} guarantee that the total gas density will remain equal to
the sum of partial densities after interpolation. One might expect
that the magnitude of this problem will be large whenever the new
patch is created in regions where the partial densities vary
significantly. Furthermore, the degree of mismatch between the total
and partial densities should decrease with increasing degree of
smoothness. The latter can efficiently be controlled using the
threshold for truncation error.  In what follows we ignore for the
moment nuclear reactions and focus on this interpolation problem by
presenting results obtained for the same initial data but varying
truncation error thresholds.

\begin{figure}[t]
\centerline{\epsfxsize=0.5\textwidth\epsffile{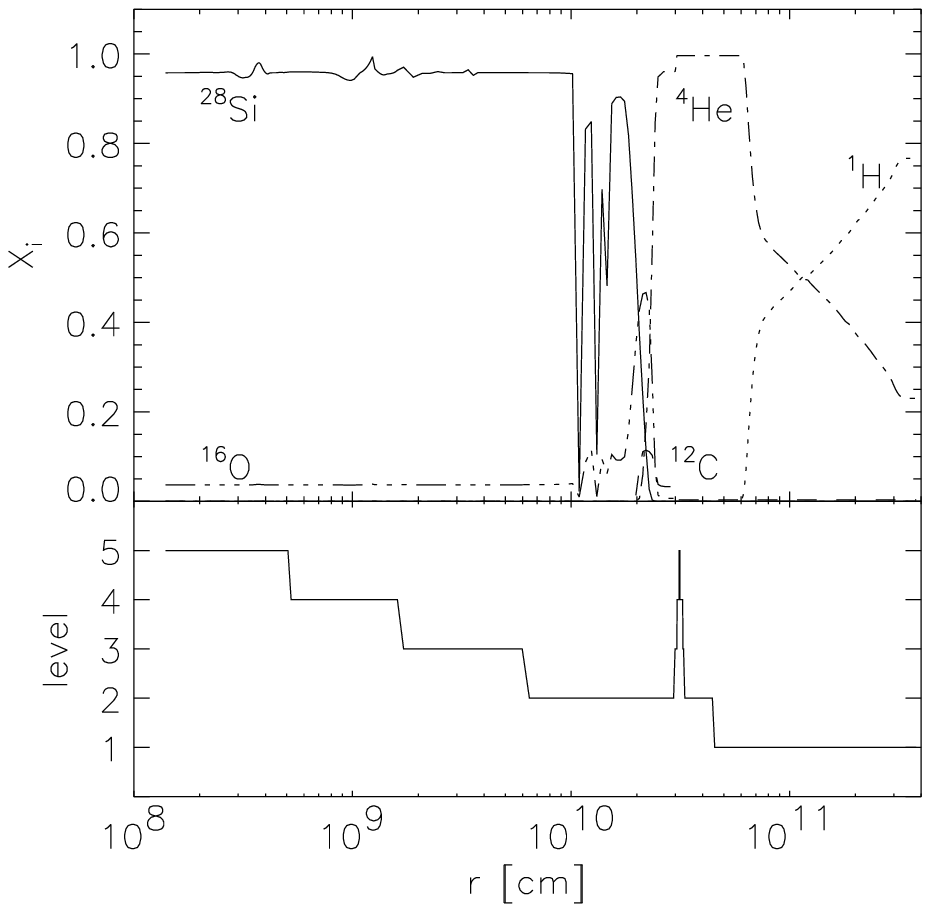}
            \epsfxsize=0.5\textwidth\epsffile{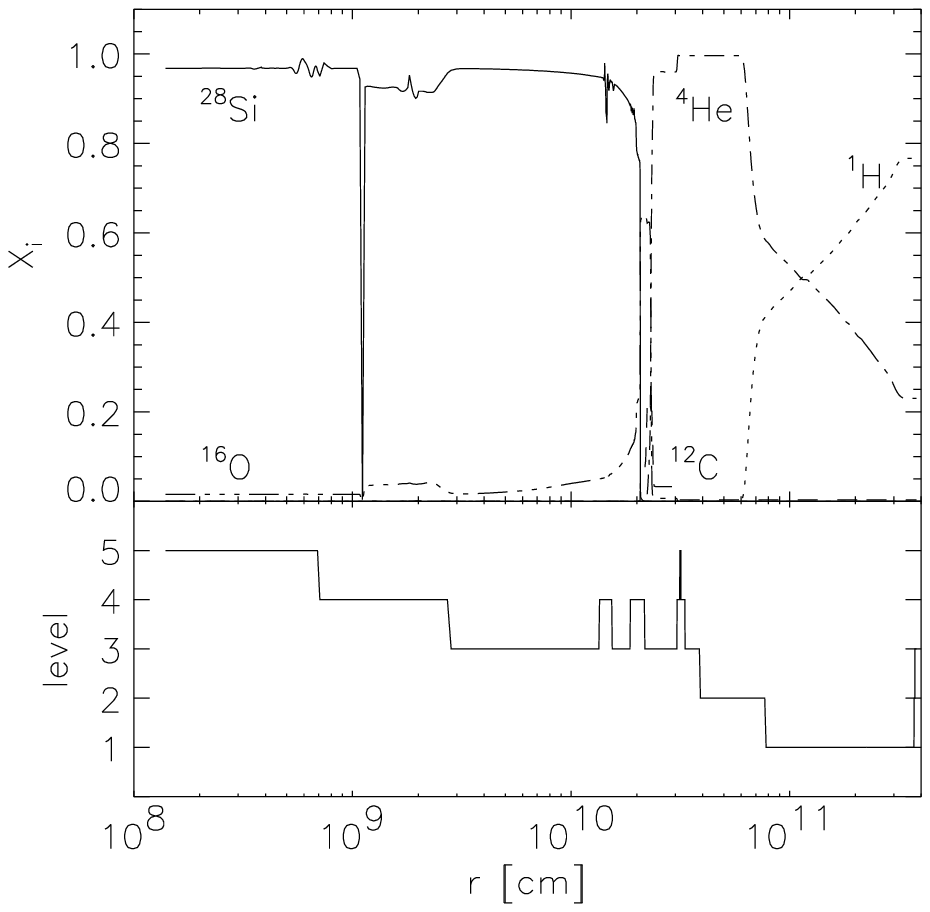}}
\caption{Left: Mass fraction profiles for our test problem
                 after 34.9\,s of evolution. By this time the 
                 shock has reached a radius slightly
                 larger than $3 \times 10^{10}$\,cm and is tracked
                 with a single level 5 patch. The error estimation
                 algorithm was applied only to
                 $(\rho, \rho u, \rho E_{\rm tot})$ and a
                 local truncation error of $\epsilon = 0.1$ was used.
                 Large errors in mass fractions can clearly be 
                 seen in the central region of the grid.
           Right: Same as left panel but with $\epsilon = 0.01$. 
                  In spite of the increased accuracy (by a factor 
                  of ten) the solution is still flawed.}
\label{Kifonidis.fig2}
\end{figure}

\begin{figure}[t]
   \centerline{\epsfxsize=0.5\textwidth\epsffile{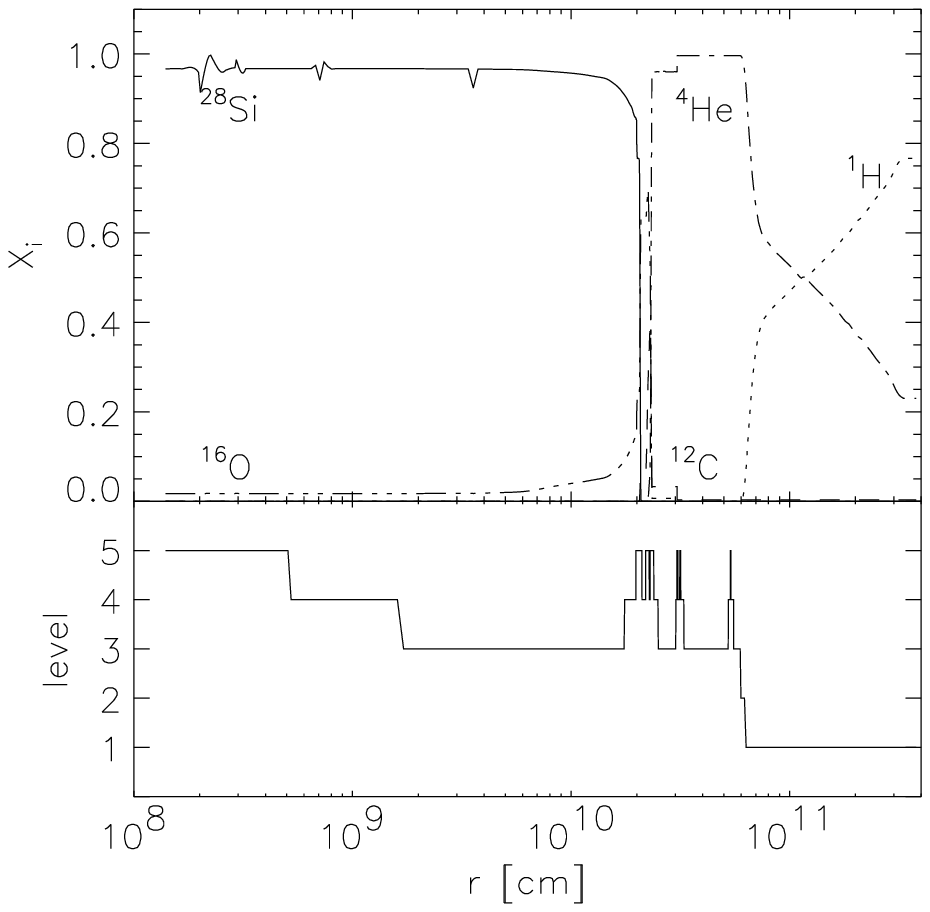}
               \epsfxsize=0.5\textwidth\epsffile{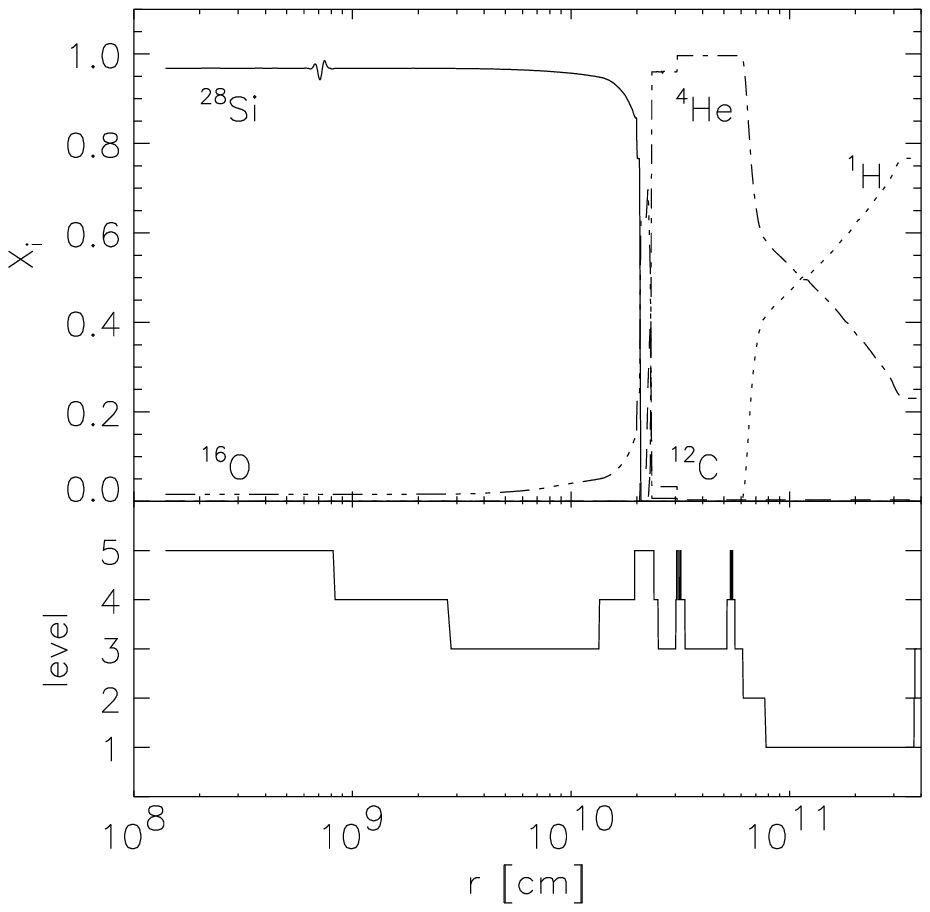}}
  \caption{Left: Same as \protect Fig.~\ref{Kifonidis.fig2}
                 but with $\epsilon = 0.1$ for
                 $(\rho, \rho u, \rho E_{\rm tot})$ 
                 and additional flagging
                 of the partial densities, $\rho X_i$, with
                 $\epsilon_X = 0.1$.
           Right: Same as left panel but with $\epsilon = 0.01$
                  and $\epsilon_X = 0.01$.}
  \label{Kifonidis.fig3}
\end{figure}

\begin{figure}[t]
\centerline{\epsfxsize=0.5\textwidth\epsffile{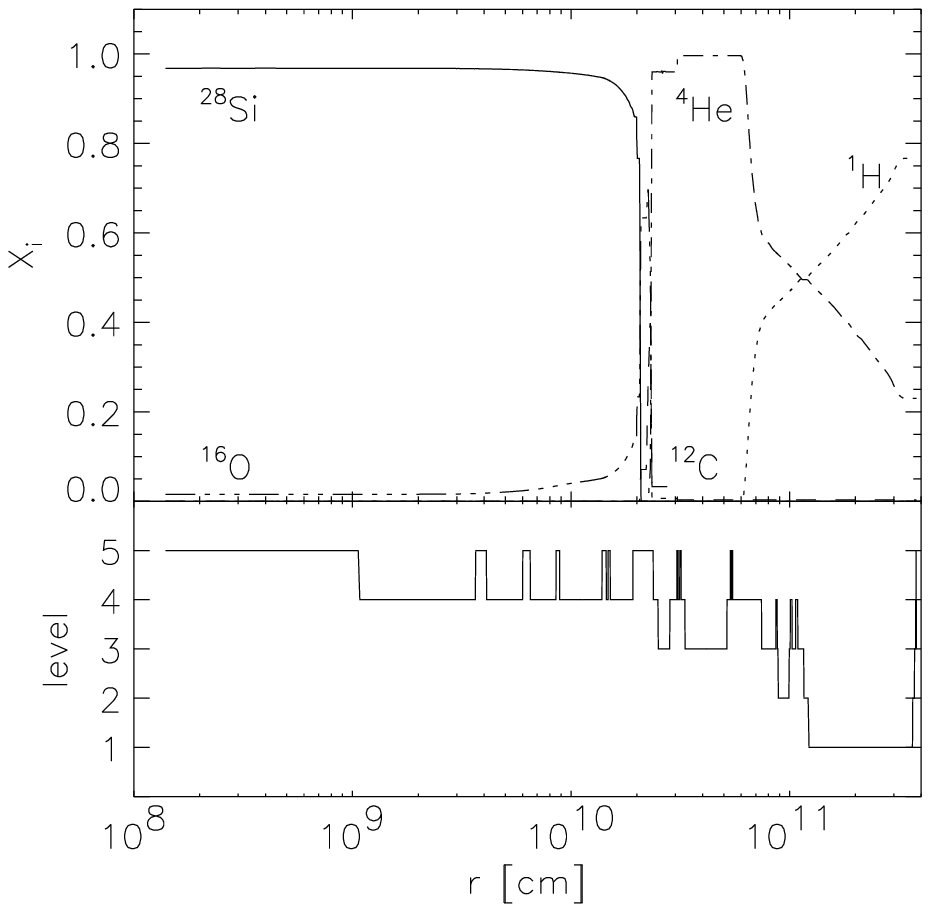}
            \epsfxsize=0.5\textwidth\epsffile{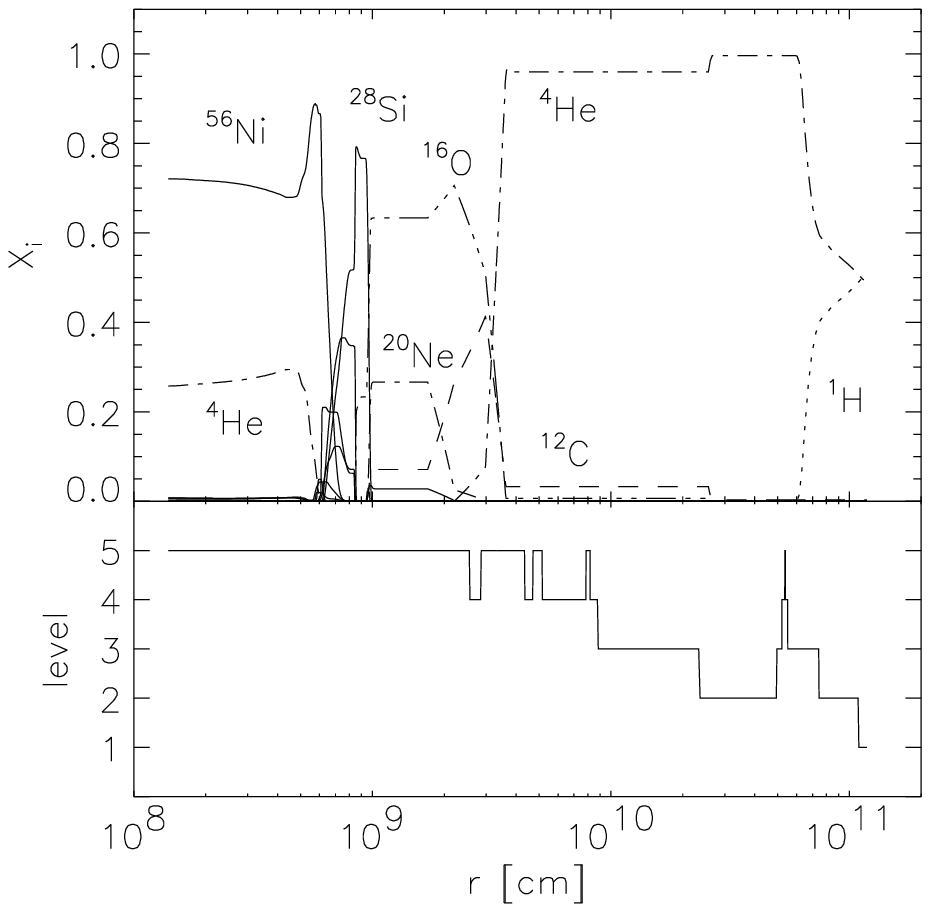}}
\caption{
Left: Same as \protect Fig.~\ref{Kifonidis.fig2} but with $\epsilon
= 10^{-3}$ and $\epsilon_X= 10^{-2}$.  All errors have
disappeared. Note the changes in the distribution of grid patches. The
larger number of level 4 and level 5 patches resulted in an
increase in CPU-time of about a factor of 5 and 3.6 as compared to the
first and second case shown in Fig.~\ref{Kifonidis.fig2},
respectively.
Right: Chemical composition at $t = 0.5$\,s for our run including
nuclear burning (see text). At this time nearly all nuclear reactions
have frozen out. Nucleosynthesis has taken place mainly in the former
silicon shell. Following the inner layers, where peak temperatures
were sufficiently high to synthesize $\rm ^{56}Ni$, incomplete
Si-burning has led to a zone dominated by $\rm ^{32}S$,
$\rm ^{36}Ar$, and $\rm ^{40}Ca$.  
The C/O-core of the star is almost completely covered with
the finest resolution ($\Delta r \approx 39$\,km). Abrupt changes in
composition in this region are a consequence of coarse zoning in the
initial model. Note that, in contrast to the other runs, the entire
grid extends up to $1.2\times 10^{6}$\,km in this case. }
\label{Kifonidis.fig4}
\end{figure}

Figure~\ref{Kifonidis.fig2} 
displays the chemical profiles obtained when the truncation error is
estimated only for the conserved quantities ($\rho$, $\rho u$, $\rho
E_{\rm tot}$) with $\epsilon = 0.1$ and $\epsilon = 0.01$ in the left
and right panel, respectively.  In both cases large errors in the
distribution of species are visible. Using a smaller $\epsilon$ helps
in resolving the outer edge of the silicon shell ($r \approx 1-2
\times 10^{10}$\,cm), but some low-amplitude noise can still be seen
at $r \approx 1.5 \times 10^{10}$\,cm.  However, the computed
distribution of $\rm ^{28}Si$ in the core does not seem to be
sensitive to this mild improvement in overall accuracy and in addition
to low-amplitude noise a conspicuous undershoot is present at $r
\approx 10^{9}$\,cm for the $\epsilon = 0.01$ case. 
The quality of the solution improves when
in addition to the error estimation for $(\rho, \rho u, \rho E_{\rm
tot})$ we also estimate the truncation error for the partial densities
(Fig.~\ref{Kifonidis.fig3}).
With $\epsilon = 0.1$ and $\epsilon_X = 0.1$ most of the material
interfaces located just below the helium shell are resolved and no
large errors in the silicon distribution are present. With increased
accuracy ($\epsilon = 0.01$, $\epsilon_X = 0.01$) the finest
level patches extend from the centre of the grid further out and help
in keeping the chemical composition smooth. The outer edge of the
silicon core is now covered with level 4 patches and all chemical
discontinuities are modelled using the highest resolution. However,
the errors are not totally eliminated.  The silicon abundance is still
affected near $r \approx 7 \times 10^{8}$\,cm. From our numerical
experiments we found that using $\epsilon = 0.001$ and $\epsilon_X =
0.01$ finally eliminates the problem (cf. the left panel of
Fig.~\ref{Kifonidis.fig4}) with patches on the finest level now
extending from the inner boundary to radii slightly above $r \approx
10^{9}$\,cm.

In the right panel of Fig.~\ref{Kifonidis.fig4} we finally present
results obtained with an $\alpha$-chain network of 27 reactions for
our 13 $\alpha$-nuclei.  The network was coupled to the hydrodynamics
as described in \cite{Kifonidis.Mueller}. The same explosion energy as
for the other runs was also adopted for this setup. However, the
computational domain extended from $r = 1.4 \times 10^{8}$\,cm to $r =
1.2 \times 10^{11}$\,cm. Five levels of refinement, 120 zones on the
base grid and refinement factors of 2, 4, 4 and 8 were used.  The
truncation error thresholds were set to $\epsilon = 0.001$ and
$\epsilon_X = 0.01$. In addition flagging of density contrasts above
0.1 was employed. The obtained solution does not differ from a
corresponding single grid model computed using 30\,720 equidistant
zones and demonstrates that with a cautious use of the AMR technique
it is possible to obtain physically correct results. Moreover, the
speedup achieved in calculating the first $6.4\times 10^{-2}$\,s of
evolution as compared to the single grid run amounted to a factor of
8.4 on a single node of an IBM SP2.  We note here that there is some
overhead associated with AMR because the source terms have to be
computed also in the error estimation procedures. This is especially
important during this early phase, when the solution of the nuclear
network dominates the computational time.  But since cooling due to
the strong expansion leads to a rapid freezeout of nuclear reactions,
we may expect AMR to offer even larger savings in CPU time during the
late evolutionary phases.  We were not able to continue this
comparison further in time, however, as the computational cost for the
single grid run turned out to be prohibitively high.

In the future, we plan to use {\sc amra} to study the problem of
nucleosynthesis and mixing in two dimensions starting shortly after
shock stagnation, when shock revival due to neutrino heating and
convective motion begins, through the stage where the aspherical shock
overruns the Si and O shells leading to an aspherical distribution of
newly synthesized nuclei, up to the development of the Rayleigh-Taylor
instability.  Current multidimensional simulations of the delayed
explosion mechanism (cf. \cite{Kifonidis.HBHFC}, \cite{Kifonidis.BHF},
\cite{Kifonidis.JM}) indicate that explosive burning will partly
proceed for electron fractions well below $Y_e \approx 0.5$ and thus
results in neutron rich isotopes.  In order to avoid a contamination
of the interstellar medium with the wrong nucleosynthetic products,
fallback of this material onto the central remnant in the late stages
of the explosion was suggested. Therefore, another goal of such
computations is to determine the actual location of the mass cut and
to provide the link needed to test the current ideas behind the
delayed explosion mechanism by confronting the ejected nucleosynthesis
products with observations.

\subsection*{Acknowledgements}

We are very grateful to Stanford Woosley for providing us with 
various progenitor models which we have used to construct
our initial data. The work of TP was partly supported by the grant KBN
2.P03D.004.13 from the Polish Committee for Scientific Research. The
simulations were performed on the IBM SP2-P2SC and CRAY J916/16512 
located at the Rechenzentrum Garching.


\end{document}